# DUAL-DISK GALAXIES AND THERMAL STATES OF CIRCUMGALACTIC MEDIUM

Masafumi Noguchi [1]
[1]Astronomical Institute, Tohoku University, Aoba-ku, Sendai 980-8578, Japan

## ABSTRACT

Disk galaxies often have dual-disk structures composed of thin disks and thick disks. Recent observations reveal that such duality is ubiquitous across broad galaxy mass and cosmic time. It is suggested that more massive galaxies have smaller mass fractions of thick disks and undergo a transition from thick (or single) disk states to dual-disk states at earlier times. Previous studies suggest that the circumgalactic medium (CGM) in halos above the critical mass is heated by the development of stable shocks and feedback from active galactic nuclei. It has been argued that this thermal change of the CGM causes the transition of the accretion of CGM from the fast cold mode to the slow hot mode driven by the radiative cooling of heated CGM. We consider here the hypothesis that the cold and hot modes promote the formation of thick and thin disks, respectively. Previous theoretical studies showed that this hypothesis explains the age and chemical bimodality of the Milky Way disk as well as the mass dependence of disk duality observed for local galaxies. Using simple galaxy evolution models, we show that this hypothesis also reproduces the thick-to-thin disk mass ratios observed for high-redshift galaxies up to $z = 2$. Earlier transition to dual disks in more massive galaxies is explained as the outcome of earlier crossing of the critical halo mass by more massive halos. The caveat here is that this interpretation is not applicable to bulge-dominated galaxies at the high-mass end because present models do not include galaxy mergers, which likely act as the primary route to bulge formation. It should also be noted that the link between the disk duality and gas accretion modes may be indirect and multiple processes may transform accreted material into different types of disks in real galaxies. Past theoretical studies suggest that the mode change in CGM accretion underlies earlier and stronger quenching of star formation in more massive galaxies and the increasing bend with time of the star-forming galaxy main sequence at the high-mass end. We infer that the change in the CGM thermal state is inscribed in diverse characteristics of galaxies.

## 1. INTRODUCTION

Complex structures of disk galaxies pose a tough challenge to galaxy formation theory. The existence of distinct disk and bulge components and their relative dominance brings about a remarkable diversity in galaxy morphology, forming the basis for the traditional Hubble classification (e.g. Simard et al. 2011; Mendel et al. 2014). Disks sometimes exhibit a dual structure comprising thin and thick disks with different scale heights (e.g. Burstein 1979; Tsikoudi 1980; Comerón et al. 2014, 2018). It is suggested that the Milky Way also has two distinct disk components with different chemical compositions and spatial extents (e.g. Haywood et al. 2013; Hayden et al. 2015; Pouliasis et al. 2017), although some doubt has been cast on the claimed bimodality (Bovy et al. 2012). Observations of local galaxies show that low-mass galaxies generally contain significant thick disks but their contribution decreases for higher mass galaxies (e.g. Yoachim & Dalcanton 2006; Comerón et al. 2014). The recent study by Tsukui et al. (2025) using JWST has extended the analysis of dual-disk structures to a wide domain of mass and redshift and made it possible to examine the emergence of dual-disk galaxies in a cosmological perspective.

A wide variety of formation mechanisms have been proposed for the thick-thin disk duality (see discussion in Funakoshi et al. 2025). They include radial migration of kinematically hot stars from inner galactic regions (e.g. Minchev et al. 2013), the stirring-up of the existing thin disks by mergers with smaller galaxies (e.g. Quinn et al. 1993; Thulasidharan et al. 2024), and conversion of merging galaxies to thick disk components (e.g. Yoachim & Dalcanton 2006; Pinna et al. 2019). However, it is not clear whether these mechanisms can explain the mass and redshift dependence of disk duality as observed.

Inspired by the theoretical framework of cold/hot accretion (e.g. Kereš et al. 2005; Dekel & Birnboim 2006), Noguchi (2020) associated different modes of gas accretion from halos to different stellar components of galaxies and examined the formation of those components. This approach has led to a reasonable reproduction of the relative contributions of thin disks, thick disks, and bulges in local galaxies (Comerón et al. 2014). Here, following Noguchi (2020), we propose that the emergence of dual-disk galaxies is closely related to the change in the thermal state of the circumgalactic medium and the resulting change in the CGM accretion mode. Such change is considered to be caused by heating by shock waves (e.g. Rees & Ostriker 1977; Kauffmann et al. 1993) and the feedback from active galactic nuclei (AGNs) (e.g. Kereš et al. 2009; Lu et al. 2014; Powell et al. 2017; Weinberger et al. 2018). As we see, this hypothesis naturally explains the overall aspect of observed dual-disk structures across the cosmic time (Tsukui et al. 2025).

Section 2 briefly describes important physical processes occurring in circumgalactic regions with particular em-

noguchi@astr.tohoku.ac.jp

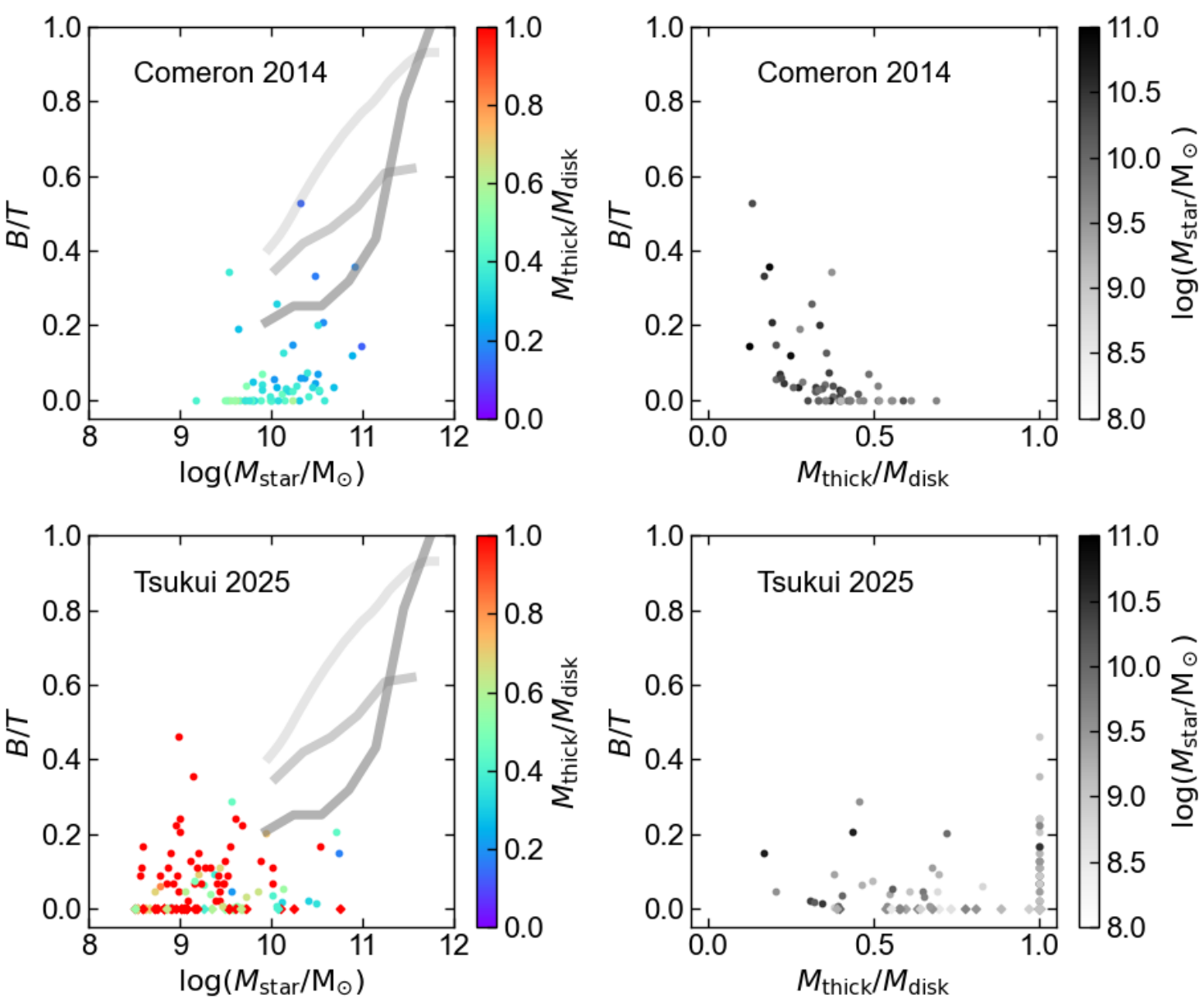

Fig. 1.— The bulge-to-total stellar mass ratios plotted against the total galaxy stellar mass (left panels) and the thick-to-total disk mass ratio (right panels). The samples of Comerón et al. (2014) and Tsukui et al. (2025) are shown in upper and lower panels, respectively. In left panels, galaxies are color-coded in accordance with the thick disk mass fraction, while gray colors in right panels indicate the total stellar mass of individual galaxies.

phasis on the cold mode accretion. We also describe past attempts to connect CGM physics to galaxy formation which are relevant to the present study. Section 3 describes the galaxy evolution code, mainly focusing on the modification of the previous one made specially for the present purpose. Results concerning dual-disk structures and comparison with observations are given in section 4. We discuss more general baryonic properties of the model galaxies in section 5. Discussion and conclusions are given in sections 6 and 7, respectively. APPENDIX describes a supplementary analysis including galaxies possessing bulges.

## 2. CGM PHYSICS AND GALAXY FORMATION

Galaxy formation process is greatly controlled by the inflow of gas from the circumgalactic space, which is roughly defined from the outer boundary of galaxies to the virial radius of their parent halo. Recent studies revealed that the circumgalactic medium generally has a complicated structure comprising multiple gaseous constituents with different physical states (e.g. Tumlinson et al. 2017)

Theoretical attempts to understand the physics of circumgalactic gas in relation to galaxy formation have a long history. The long-standing paradigm, the shock-heating theory (e.g. Silk 1977; Rees & Ostriker 1977; White & Rees 1978), argues that the gas that entered the halo is heated by shock waves to high (virial) temperatures, and then gradually accretes to the central part as it cools by emitting radiation (cooling flow). This picture predicts a smooth and continuous gas accretion history. Therefore, it is not straightforward to understand why and how the different components of disk galaxies formed.

The shock-heating theory was later modified by the cold-accretion scenario, which proposes that the shock waves do not arise in some cases and a significant part of the gas stays unheated, reaching the forming galaxy in narrow streams almost in a free-fall fashion (cold accretion) (e.g. Fardal et al. 2001; Birnboim & Dekel 2003; Kereš et al. 2005; Dekel & Birnboim 2006). The hypothesis of cold accretion was found to explain the large abundance of massive star-forming galaxies at high redshifts, which is difficult to explain by the simple shock-heating picture (Dekel et al. 2009). It is also invoked to quench star formation in massive galaxies in cooperation with the feedback from active galactic nuclei in theoretical models (e.g. Croton et al. 2006). Using the boundary between hot and cold accretion modes in the redshift-halo mass plane predicted by Dekel & Birnboim (2006), Noguchi (2023) reproduced the star formation quenching in massive galaxies at high redshift (e.g. Behroozi et al. 2013, 2019; Moster et al. 2018) as well as the time evolution of the star-forming main sequence of galaxies (e.g. Whitaker et al. 2014; Tomczak et al. 2016), i.e., increasingly strong bending at high galaxy masses.

The predicted cold gas streams are claimed to be found in the form of quasar absorption line systems such as Lyman limit systems (Fumagalli et al. 2011) or Lyman $\alpha$ emitting filaments around forming galaxies (Martin et al. 2016). Theoretically, early cosmological simulations mainly using the Smoothed Particle hydrodynamics (SPH) showed the clearcut emergence of filamentary structures of inflowing cold halo gas (e.g. Kereš et al. 2005; Ocvirk et al. 2008; van de Voort & Schaye 2012). These results were, however, sometimes criticized as numerical artefacts inherent in SPH by later simulations using mesh-based codes (e.g. Nelson et al. 2013).

Noguchi (2018) applied the cold accretion scenario to the Milky Way (MW), and showed that the two chemically different stellar groups observed in the MW disk (e.g. Haywood et al. 2013; Hayden et al. 2015) naturally arise. In this picture, the stellar group rich in $\alpha$ elements (O, Mg, Si, etc) was formed by a fast star formation event fueled by the cold accretion in early times whereas the $\alpha$-poor stellar group was gradually formed by the cooling flow accretion of the shock-heated gas in later epochs. Interestingly, the $\alpha$-rich stars in the solar neighborhood tend to have larger velocities perpendicular to the disk plane than $\alpha$-poor stars (e.g. Haywood et al. 2013), meaning that they reach higher altitudes resulting in a thicker configuration. Although this correspondence is not always clear-cut (Haywood et al. 2013), the result of Noguchi (2018) hints at the association of the cold accretion and the cooling flow with the formation of the MW thick disk and thin disk, respectively (also see Brooks et al. 2009; Hafen et al. 2022).

This dual-disk structure is not specific to the Milky Way. Local galaxies generally possess two disk components in addition to bulges (Comerón et al. 2014). Noguchi (2020), building on the correspondence between different accretion modes and different stellar components suggested in the above MW study, reproduced the decreasing mass fraction of thick disks and increasing bulge mass fraction toward higher-mass galaxies in the local galaxy sample. The observational analysis by Tsukui et al. (2025) using JWST enables a further test of the hypothesis made by Noguchi (2018) thanks to its extended mass and time domain covered by the sample. In this study, we use the galaxy formation model including cosmologically-motivated accretion schemes for the CGM and confront the model results with observational data.

## 3. GALAXY EVOLUTION MODEL

Our method, like the semi-analytic models (SAM) (e.g. Lacey et al. 2016; Croton et al. 2016; Mitchell et al. 2018) and the equilibrium models (e.g. Davé et al. 2012; Mitra et al. 2015), does not rely on numerical simulations but traces galaxy evolution analytically except introducing simulation results as input physics. Unlike SAM, we do not address the population properties such as galaxy mass functions, luminosity functions, and spatial correlation functions. We instead follow the galaxy evolution for various halo masses specified at the present epoch by time integration from the appropriate initial condition. In addition to the mass growth of dark matter halos, we analytically model fundamental baryonic processes such as accretion of halo gas to forming galaxies, star formation from accreted interstellar medium (ISM), ejection of ISM by feedback processes, and recycling of the ejected ISM. We permit non-equilibrium states of the system unlike the equilibrium approach. For example, the mass of the interstellar medium is determined by the balance of the replenishment by accretion from halos and the consumption by star formation and therefore varies with time generally. Our modelling is admittedly idealized, missing many details included in cosmological simulations and SAM. This approach, however, helps grasp the effect of each physical process without delving into complexities of galaxy evolution.

Detailed description of the model calculation is given in a previous series of papers (Noguchi 2020, 2021, 2022, 2023), which investigate the evolution of galaxies in various accretion schemes of the halo gas. Here we briefly summarize the essential ingredients of the model. The starting point is the time evolution of the halo mass for a given value of the present mass. We use the parametrization of the mass growth history given in APPENDIX H of Behroozi et al. (2013), which fits very well with the results of Bolshoi (Klypin et al. 2011), MultiDark (Klypin et al. 2016), and Consuelo cosmological dark matter simulations. Once the halo growth history is determined, we calculate the mass accretion rate of gas into halos by assuming the universal baryon fraction of $f_{\rm b} = 0.17$. We assume that when the mass of a halo increases by $\Delta M_{\rm halo}$, its portion $f_{\rm b}\Delta M_{\rm halo}$ is attributed to the increment of the halo gas.

We next calculate the accretion rate of the halo gas onto the galaxy. In the most general form, the cold-accretion theory predicts three regimes for the thermal state and accretion process of the CGM depending on the halo mass, $M_{\rm halo}$, and redshift $z$ (e.g. Kereš et al. 2005; Dekel & Birnboim 2006; Ocvirk et al. 2008). In this paper, we refer to the virial mass of a halo as the halo mass. There are two important mass scales in defining those regimes. First, $M_{\rm shock}$ is the halo mass above which a stable shock develops and the CGM is heated to the virial temperature. Second, $M_{\rm stream}$ is the mass above which the narrow inflowing streams of cold unshocked gas disappear. Therefore, the cold accretion dominates when $M_{\rm halo} < M_{\rm shock}$ (Domain F), whereas the cooling flow dominates in Domain G ($M_{\rm halo} > M_{\rm shock}$ and $M_{\rm halo} > M_{\rm stream}$). The halo gas in Domain H ($M_{\rm shock} < M_{\rm halo} < M_{\rm stream}$) has a unique hybrid structure with narrow streams of the cold gas penetrating ambient shock-heated gas. We assume that the cold gas and the shocked gas experience the cold accretion and the cooling flow, respectively. The fate of the cold gas in Domain H is not clear. Noguchi (2020) tentatively associated this component with galactic bulges because such a treatment leads to the increasing dominance of bulges toward more massive galaxies, as observed.

Here, we assume that thick and thin disks are formed by the cold and hot modes of gas accretion, respectively, on the basis of the insight into the origin of those disks in the Milky Way (Noguchi 2018). We further assume that the accretion timescale of the cold and hot modes is the free-fall time and the radiative cooling time, respectively. The former is simply determined by the mean mass density of the halo, which is in turn determined by the cosmological mean density for the specified cosmology. The latter is calculated from the density, temperature and metallicity of the halo gas, using the table for the radiative cooling time under the collisional ionization equilibrium given in Sutherland & Dopita (1993).

We briefly describe the treatment of baryonic processes associated with the accreted gas. See Noguchi (2023) for more details. We follow Blitz & Rosolowsky (2006) in implementing the star formation process. The gas accreted from the halo is assumed to form a gaseous disk and replenish the interstellar medium (ISM), from which stars are formed. Star formation rate density, $\Sigma_{\rm SFR}$, is given by

$$\frac{\Sigma_{\rm SFR}}{({\rm M_\odot pc^{-2} Gyr^{-1}})} = \frac{\epsilon_{\rm SF}}{({\rm Gyr^{-1}})} \frac{\Sigma_{\rm HCN}}{({\rm M_\odot pc^{-2}})} \quad (1)$$

where $\epsilon_{\rm SF} = 13{\rm Gyr^{-1}}$ (Gao & Solomon 2004; Wu et al.

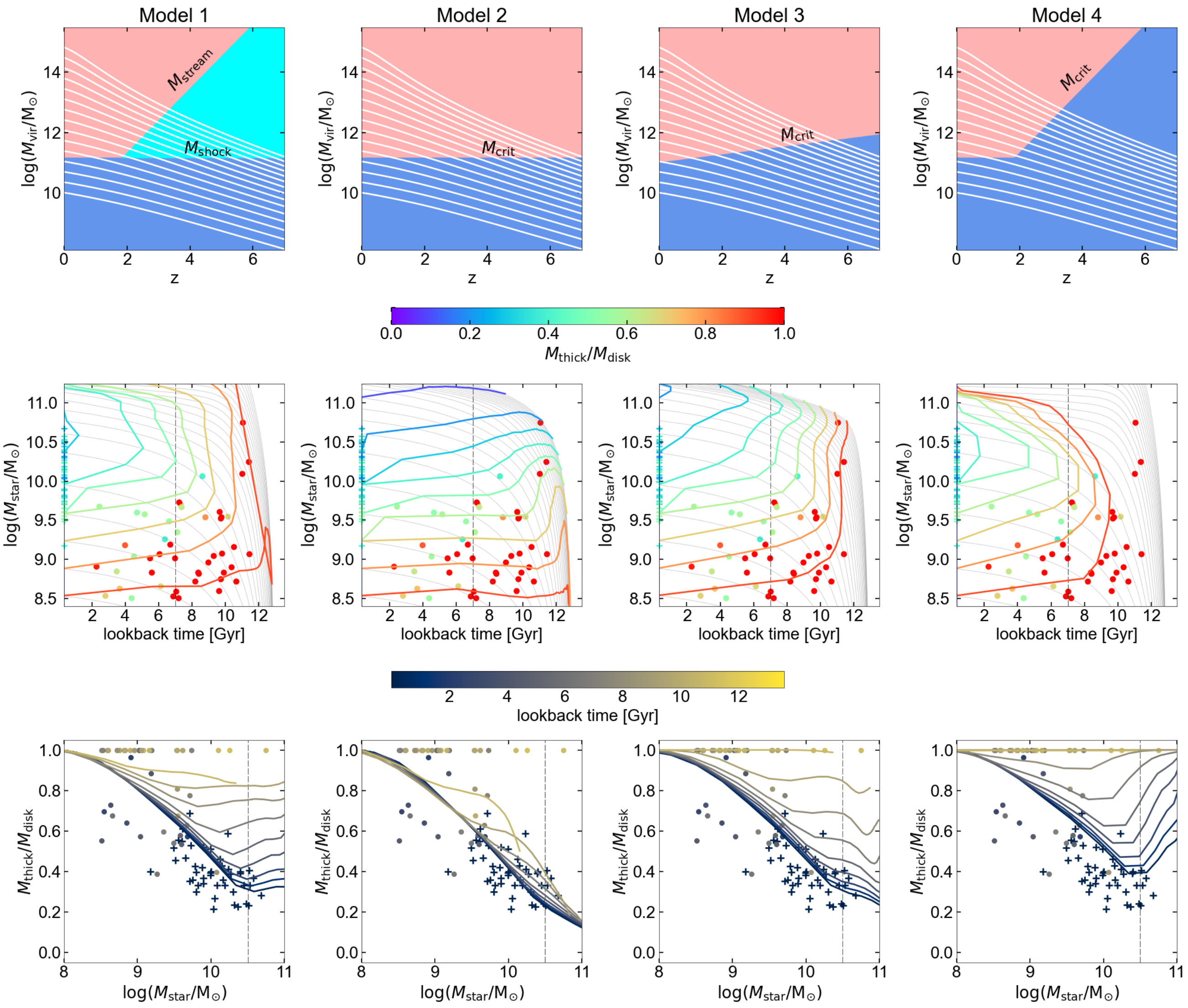

FIG. 2.— Time evolution of the mass fraction of thick disks with respect to the total mass of stellar disks. Top panels : Domains for cold (blue) and hot (red) mode accretion for four calculated models. $M_{\rm vir}$ is the halo mass and $z$ stands for redshift. The cyan region in Model 1, bounded by the shock mass, $M_{\rm shock}$, and the stream mass, $M_{\rm stream}$, is Domain H, in which cold and hot modes coexist. In Models 2, 3, and 4, the domains for cold and hot modes are divided by the critical halo mass, $M_{\rm crit}$. White lines in each panel indicate the growth of halo mass for selected model galaxies. Middle panels : Colored contours show the thick disk fraction as a function of the galaxy total stellar mass and the lookback time. Contour levels are taken from 0.1 to 0.9 equally spaced by 0.1. Gray lines indicate stellar mass growth of model galaxies. The samples of Comerón et al. (2014) and Tsukui et al. (2025) are denoted by crosses and circles colored by thick disk fraction, respectively. Vertical dashed lines mark the epoch which divides two age ranges used in Fig.3. Bottom panels : Thick disk fraction against the galaxy mass for selected lookback times from 0 to 12.6 Gyr equally spaced by 1.4 Gyr. Observed galaxies (crosses for Comerón et al. (2014) and circles for Tsukui et al. (2025)) are colored by the lookback times. Galaxies more massive than the vertical dashed lines are mentioned 'massive galaxies' in the text.

2005) and $\Sigma_{\rm HCN}$ stands for the surface density of the dense molecular gas. Massive stars provide supernova feedback to the surrounding ISM and eject a part of it. The ejected mass is calculated as

$$\Delta M_{\rm ej} = \left[\frac{2\epsilon_{\rm FB} E_{\rm SN}\eta_{\rm SN}}{V_e^2}\right]\Delta M_{\rm star} \quad (2)$$

where $\Delta M_{\rm star}$ is the mass of stars formed in the current time step, $V_e$ is the escape velocity of the halo, $\epsilon_{\rm FB}$ is the feedback efficiency, $E_{\rm SN}$ is the energy released by one supernova, and $\eta_{\rm SM}$ is the number of supernovae produced per one solar mass of stars formed at that step. We take $E_{\rm SN} = 10^{51}$ erg and $\eta_{\rm SN} = 8.3\times10^{-3}$ following Dutton & van den Bosch (2009). For feedback efficiency we adopt $\epsilon_{\rm FB} = 0.2$ as the fiducial one.

The ultimate fate of the ejected gas is highly uncertain. We assume that some portion of the ejected gas falls back and returns to the galactic disk for further star formation (e.g. Oppenheimer et al. 2010; Henriques et al. 2013; Christensen et al. 2016). We use the parametrization by Mitra et al. (2015) for the timescale of reincorporation (recycling) of the ejected gas as

$$t_{\rm rec} = 0.52\times10^9{\rm yr}\times(1+z)^{-0.32}\left(\frac{M_{\rm halo}}{10^{12}{\rm M}_\odot}\right)^{-0.45} \quad (3)$$

This is obtained as the best fit for their equilibrium model. The mass recycled at each time step is calculated as

$$M_{\rm rec} = (M_{\rm ej}/t_{\rm rec})\,\Delta t \quad (4)$$

where $M_{\rm ej}$ is the total mass of ejected material present at the current time step (i.e., the reservoir of the ejected material) and $\Delta t$ is the (constant) length of a single time step. It should be noted that $M_{\rm ej}$ increases by the stellar ejective feedback and decreases by recycling. Therefore, it is updated at each time step as

$$M_{\rm ej} = M_{\rm ej} + \Delta M_{\rm ej} - M_{\rm rec} \quad (5)$$

In low-mass galaxies, the gas accretion is assumed to be suppressed by the preventive feedback accompanying galactic winds driven by the supernova feedback. This type of feedback was introduced to solve a tension between mass and metallicity in local low-mass galaxies in semi-analytic models (e.g. Lu et al. 2017). Cosmological simulations show that outflowing winds indeed cause the preventive feedback (van de Voort et al. 2011; Mitchell et al. 2020; Mitchell & Schaye 2022). To mimic the preventive feedback in a simple form, we reduce the halo gas accretion rate to the disk calculated as above to the level

$$f_{\rm p} = f_{\rm pr} + (1 - f_{\rm pr})(y + \frac{\pi}{2})$$

, where

$$y = \arctan[3(\log M_{\rm halo} - \log M_{\rm pr})/w_{\rm pr}]$$

Here $f_{\rm pr}$ is the parameter controlling the strength of the feedback, $M_{\rm pr}$ is the critical mass below which the feedback is significant, and $w_{\rm pr}$ controls how suddenly the feedback gets strong with the decreasing halo mass. $M_{\rm pr}$ is a function of redshift and given by

$$\log M_{\rm pr} = M_0 + z(M_5 - M_0)/5,$$

where $M_0$ and $M_5$ are the critical mass at $z = 0$ and 5, respectively. We take $f_{\rm pr}$=0.3, $M_0$=11.8, $M_5$=11.0, and $w_{\rm pr}$=0.5 as the default values. For this choice, $f_{\rm p} \sim 0.65$ at the critical mass. The original accretion rate is multiplied by the calculated factor $f_{\rm p}$ to model the preventive feedback. See Noguchi (2023) for details about choice of these parameters.

One complication in the present modelling is that, generally, two gaseous components originating in cold and hot accretion co-exist in the interstellar matter (ISM) in model galaxies because conversion into stars takes finite times. We include the ejection of ISM by supernova(SN) feedback and re-accretion of ejected gas (recycling). The ejected and recycling gas also comprises two components with different origins. In model calculation, we update the mass fractions of two components in ISM every time step on the basis of cold-mode and hot-mode gas accretion rates. Thick and thin disk stars were created in accordance with their mass fractions. We assume that, at each time step, the portion of ISM ejected by supernova explosion has the same mixture of two gaseous components as in ISM and is added to the reservoir of ejected gas. The component ratio in this reservoir is updated at every timestep. The portion of gas in the reservoir that falls back to disks is assumed to have the same component ratio as in the reservoir. Detailed numerical simulations would be required to check the validity of these treatments.

In comparing model results with observational data, a caution is needed to extract appropriate samples of observed galaxies. Past observational studies have been carried out mostly on disk-dominated galaxies (Yoachim & Dalcanton 2006, 2008; Comerón et al. 2014; Tsukui et al. 2025). Fig.1 shows the characteristics of the observed galaxies which we use for model comparison in this study. It is seen that both Comeron and Tsukui data are dominated by galaxies with $B/T \lesssim 0.3$. We pick up Comeron galaxies with $B/T < 0.1$ and the pure disk galaxies in the Tsukui sample for later analysis. This restriction is made for straightforward comparison with the models. It is not clear what effects galaxy mergers have on dual-disk structures, complicating analyses including galaxies with non-negligible bulges. To be consistent with this criterion, we do not include mergers in our models and treat galaxies and their parent halos as isolated systems. In APPENDIX, we confirm that including sample galaxies with bulges for comparison does not change conclusions of this study.

We consider four variants of CGM accretion scheme as shown in top panels of Fig.2. Model 1 has the same scheme as adopted in Noguchi (2020), which succeeded in reproducing the mass fractions of thin disks, thick disks, and bulges in local galaxies. In Domain H, we assume that half the CGM accretes in the cold mode and another half in the hot mode. Because the existence of the hybrid Domain H is not firmly established, in other Models, we introduce the critical mass scale, $M_{\rm crit}$, which divides cold-mode and hot-mode accretion corresponding to unheated and heated CGM, respectively. In Models 2, 3, and 4, all of the CGM in halos above the critical mass is assumed to accrete in hot-modes. Model 2 has the critical halo mass, $M_{\rm crit}$, which stays constant with time. Such a flat critical mass is expected in the shock-heating picture, in which the cosmic gas that enters dark matter halos is always heated by shock waves to the halo virial temperature. $M_{\rm crit}$ increases mildly with redshift in Model 3. Finally, Model 4, in which Domain H in Model 1 is replaced by the completely cold domain, has a steeply increasing $M_{\rm crit}$. The cold and hot mode accretions are assumed to proceed with the free-fall time and the radiative cooling time, respectively.

We calculate the evolution of 30 model galaxies for the present halo masses equally separated logarithmically between $10^{10}{\rm M}_\odot$ and $10^{15}{\rm M}_\odot$. We adopt WMAP1 cosmology (Spergel et al. 2003) with the cosmological parameters $\Omega_{\rm M} = 0.25$, $\Omega_\Lambda = 0.75$, and $H_0 = 73{\rm kms}^{-1}{\rm Mpc}^{-1}$.

## 4. MAIN RESULTS ON DISK DUALITY

Fig. 2 confronts our models with observational data on the development of dual-disk components. Middle panels show the mass and time dependence of the thick disk mass fraction in stellar disk components. Observations show that more massive galaxies have smaller thick disk fraction at the fixed lookback time and this fraction decreases with time at the fixed stellar mass. In other words, the transition from thick-dominated to thin-dominated structure occurs earlier in galaxies with larger present halo (and stellar ) masses. We first discuss Model 1. This model shows qualitatively the same trend as ob-served except for the high mass range $M_{\rm star} > 10^{10.5}{\rm M}_\odot$. Bottom panels illustrate the model results from a different point of view. Solid lines give the thick disk fraction as a function of galaxy mass for selected lookback times. The model shows the thick disk fraction which decreases as the stellar mass for the range $M_{\rm star} < 10^{10.5}{\rm M}_\odot$, which covers most observed galaxies. This relation gets steeper toward more recent epochs, in agreement with the observational trend. The model shows a turn-up for the high-mass range, which corresponds to large thick disk fractions seen for most times in the middle panel. This transition occurs at $M_{\rm star} \sim 10^{10.5}{\rm M}_\odot$ at the present epoch and moves to lower masses toward larger lookback times. The transition mass at the present is where the bulge starts to dominate the stellar mass budget of local galaxies as seen in Fig. 1, left panels. Most disk galaxies in the observed samples indeed reside below this mass as seen in Fig. 1. Bulges are considered to form mainly from galaxies mergers neglected in this study. Models including mergers (Noguchi, in preparation) suggest that most stars formed in early massive galaxies actually end up in bulges. Therefore, we consider that this turn-up represents the upper mass below which the present model without galaxy mergers is applicable. It is encouraging that the accretion scheme in Model 1 which succeeded in reproducing the mass budgets of three components (thin disks, thick disks, bulges) in local galaxies (Noguchi 2020) also reproduces the observations at higher redshift. Models 3 and 4 show qualitatively the same trend as Model 1 and reproduce observations whereas Model 2 exhibits redshift dependence too small. Colored contours in the middle panel are almost horizontal and the thick-disk fraction in the bottom panel decreases steeply with the galaxy stellar mass already at early times. Too early transition to the hot accretion mode in Model 2 suppresses growth of thick disks in massive galaxies. It is difficult to choose the best model from Models 1, 3, and 4 by comparing the model thick-to-thin disk mass fractions with the observed ones. Nevertheless, we may conclude that we need a critical halo mass increasing with redshift to reproduce the observed behavior of thick-to-thin disk mass ratios. A caveat here is that the models in this study do not include galaxies mergers so that they are applicable only to disk-dominated galaxies.

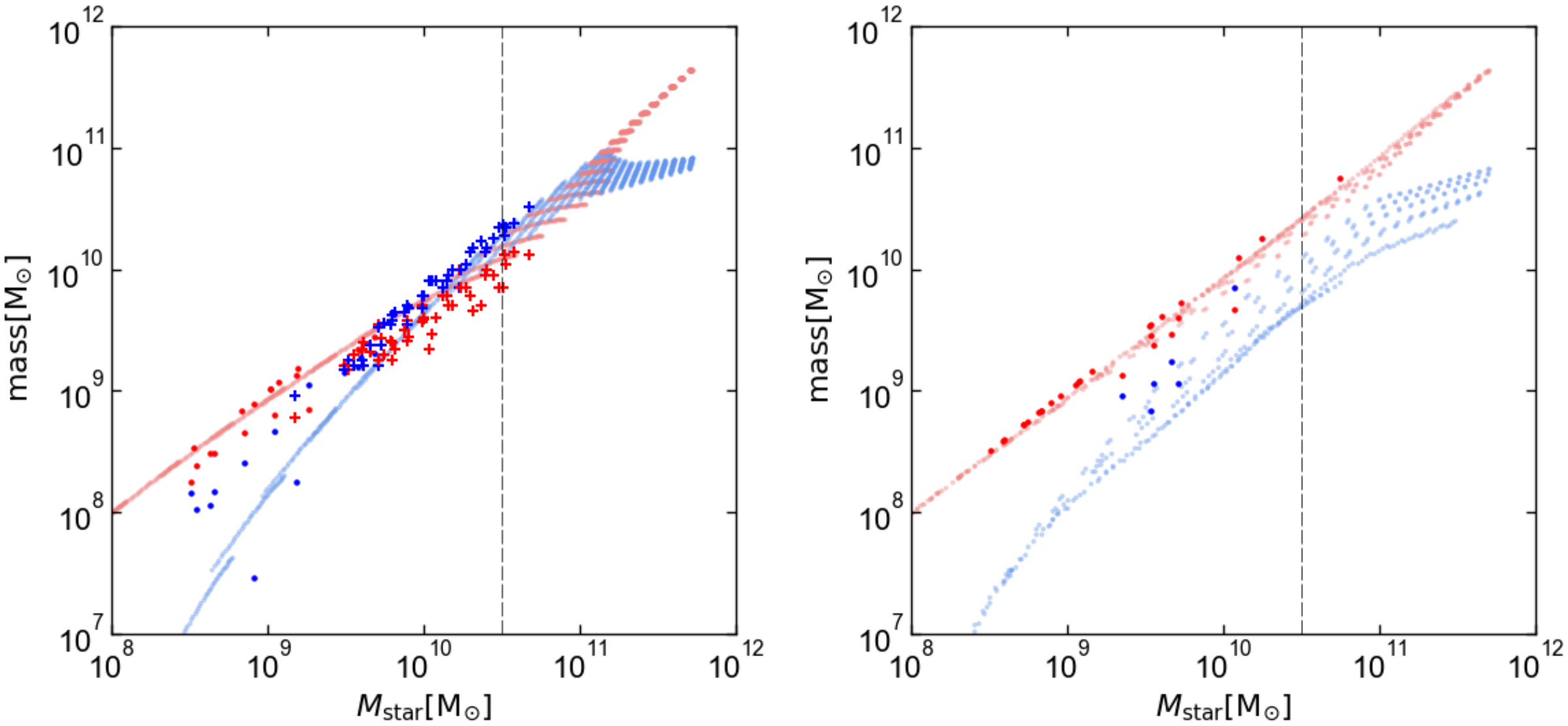

FIG. 3.— Masses of thick (red) and thin (blue) disks in Model 1 plotted against the galaxy stellar mass. Regions occupied by model galaxies are shown by shaded colors. Galaxies with $B/T < 0.1$ in Comerón et al. (2014) and purely disk galaxies (with no bulges) in Tsukui et al. (2025) are denoted by crosses and circles, respectively. Left and right panels refer to the age ranges 0-7 Gyr and 7-12 Gyr, respectively. Vertical dashed lines are the same as in the bottom panels of Fig. 2.

Fig.3 compares disk masses in Model 1 with the observational data in absolute units. Observations show a clear trend that the thin disk mass increases more steeply with the galaxy mass than the thick disk mass. Model results show a similar trend for the mass range where the observed samples are located. Both the observational and the model correlations stay roughly unchanged up to $z \sim 2$, reflecting nearly horizontal contours for constant thick disk fractions shown in the middle panel of Fig. 2. Models 2, 3, and 4 show similar results to Model 1 so that it is difficult to compare the validity of four models based on the absolute masses of thick and thin disks. Mass *ratios* of two disks provide a more powerful diagnostic. This is true for other properties of model galaxies such as disk ages, disk metallicities, baryon contents of parent galaxies. We therefore show only the result for Model 1 in the following figures to avoid unnecessary complications.

Ages of disks may provide an independent test for the model. Most observational studies on thick and thin disk ages so far concern the Milky Way, showing that its thick disk is older than the thin disk by several Gyr (e.g Haywood et al. 2016; Kilic et al. 2017). Observations for other galaxies are sparse. Fig.4 compares the ages of model disks and local galaxy disks analyzed by Yoachim & Dalcanton (2008). Observations reveal younger ages for thin disks on average despite large scatters. Thick disk ages increase for massive galaxies whereas thin disk ages stay nearly constant. Our model produces ages that run through the range occupied by local galaxies.

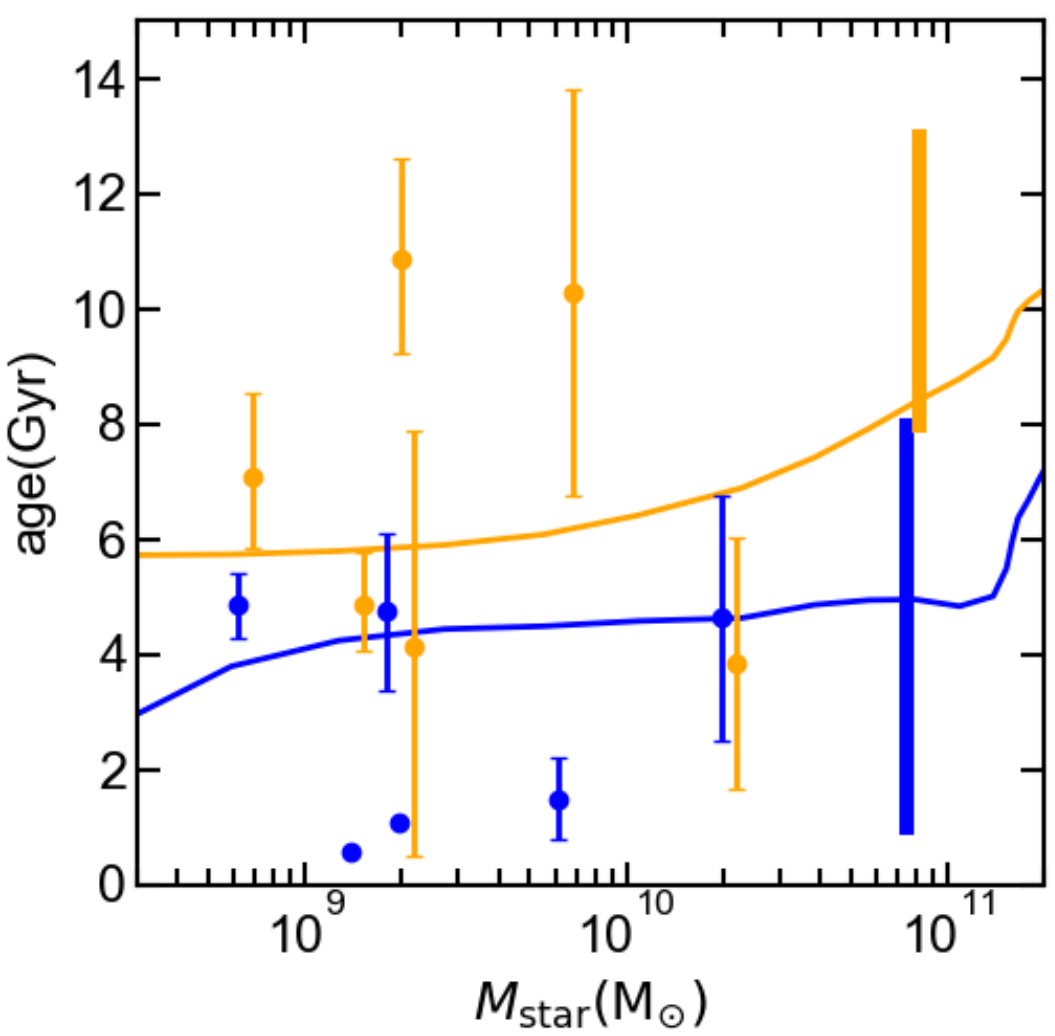

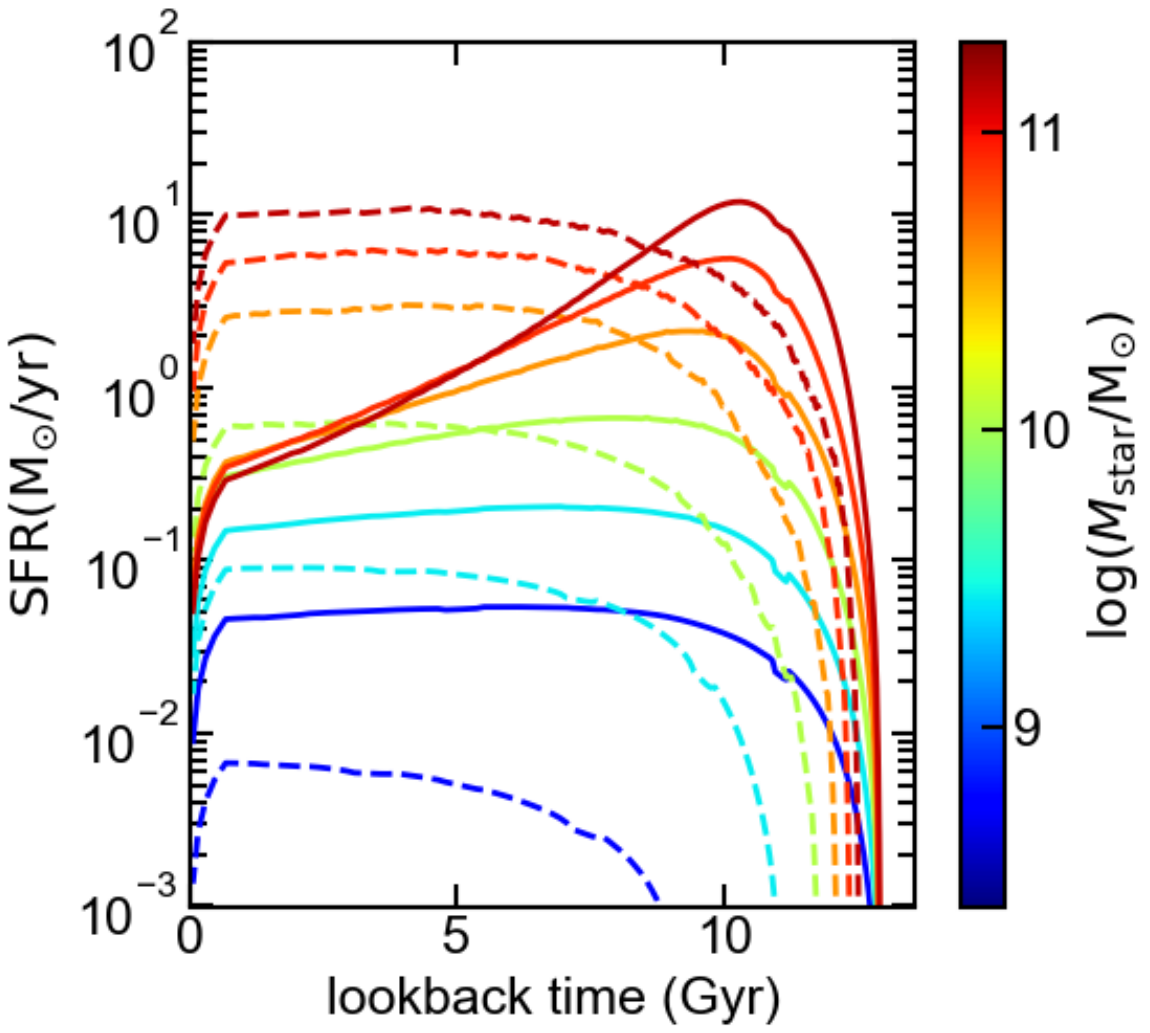

Fig. 4.— Left panel: Ages of thick (orange) and thin (blue) disks as a function of the galaxy stellar mass in Model 1. Observations by Yoachim & Dalcanton (2008) are shown by dots with error bars, with the galaxy circular velocity converted to the stellar mass using the Tully-Fisher relation by Reyes et al. (2011). Two thick vertical segments indicate the age estimate by Haywood et al. (2013) for the solar neighborhood stars. Right panel: Star formation rates for thick disks (solid lines) and thin disks (dashed lines) for selected galaxies in Model 1, with colors indicating the present stellar mass.

Right panel of Fig.4 shows the time evolution of star formation rates (SFRs) in thick and thin disks. Thick disks start to form earlier than thin disks. It is obvious that both thick and thin disks have active star formation phases spreading wide over time. This characteristic makes the concept of 'age' for these disk components somewhat ambiguous and likely causes various and sometimes discrepant age estimates for individual galaxies in the literature. Moreover, star formation histories of both disks can vary with positions in galaxies (e.g. Hayden et al. 2015; Haywood et al. 2016; Katz et al. 2021), bringing another complication. Despite these issues, it is notable that the SFR profiles for thick disks show increasingly larger drops after their peaks with the present stellar mass whereas the thin disk SFRs behave relatively similarly for the investigated mass range. This is consistent with the different mass-dependence of thick and thin disk ages in Model 1 shown in the left panel. Age estimates for a larger sample of galaxies combined with structural decomposition of thick and thin disks such as done by Yoachim & Dalcanton (2006) may provide a powerful independent test for the present model and the hypothesis behind it.

## 5. OTHER ASPECTS OF THE MODEL

We examine here how well the present model performs regarding properties other than disk duality. Upper panels of Fig.5 illustrate the mass ratio of galaxies and parent halos as a function of the halo mass. Here, the mass of galaxies includes contributions from stars and ISM. Most information about neutral hydrogen at high redshift is currently obtained from observations of damped Lyman absorption systems (DLAs). It is much debated whether neutral hydrogens observed as DLAs are concentrated in galactic disks or distributed in more extended regions (e.g. Villaescusa-Navarro et al. 2018; Walter et al. 2020; Gurman et al. 2025). We assume here that the neutral hydrogens in high-redshift galaxies are distributed predominantly in galaxy disks as a part of ISM as seen in local galaxies. Solid lines indicate the result of empirical modelling by Popping et al. (2015) (left panel) and Guo et al. (2023) (right panel). Both empirical analyses indicate the peak mass ratio at $M_{\rm halo} \sim 10^{12} M_\odot$ for $0 < z < 2$. Our model provides overall agreement with empirical ones. The lower left panel plots the ratio of the galaxy stellar mass and the parent halo mass against the halo mass. The empirical result by Behroozi et al. (2013) shows the peak at $M_{\rm halo} \sim 10^{12} M_\odot$ for $0 < z < 3$. The model seems to overproduce stars at $2 < z < 3$ though it gives an excellent match at $z < 1$. The discrepancy seen here is probably due partly to the idealized treatments of physical processes in the evolution code used here. It should also be borne in mind that the observation at high redshift (especially for cold gas components) suffers from large uncertainty.

The lower right panel in Fig.5 compares the mass of molecular hydrogens. The observational data used here come from the compilation by Tacconi et al. (2020). Molecular fraction is higher at earlier times at fixed stellar masses both in observations and the model. The model, however, shows milder time evolution of the molecular mass than the observations suggest. The semi-analytic model of Popping et al. (2014), shown by dashed lines, exhibits even slower evolution. Even though, the three sets of material indicate qualitatively the same behavior of molecular contents in galaxies.

## 6. DISCUSSION

In this study, we sought the origin of the dual structure of galactic disks in the thermal states of the circumgalactic medium distributed in the extended space of the dark matter halos. Our success in reproducing the disk duality for various galaxy masses and cosmological epochs depends on the assumption that the cold and hot modes of the CGM accretion promote the formation of thick and thin disks, respectively. One reason behind this assumption is that it reproduces the mass fractions of thin and thick disks and bulges in local (i.e., present-day) galaxies as indicated in Noguchi (2020). We extended the applicability of this assumption to early cosmological epochs by confronting the model results with the advanced observational data. Brooks et al. (2009) investigated the accretion history of the different components of the numerically simulated disk galaxies and showed that the shock heated gas tends to end-up in thin disks, while the thick disks and bulges are predominantly formed from the gas accreted in unheated states. The present model is thus in broad agreement with the cosmological simulation by Brooks et al. (2009).

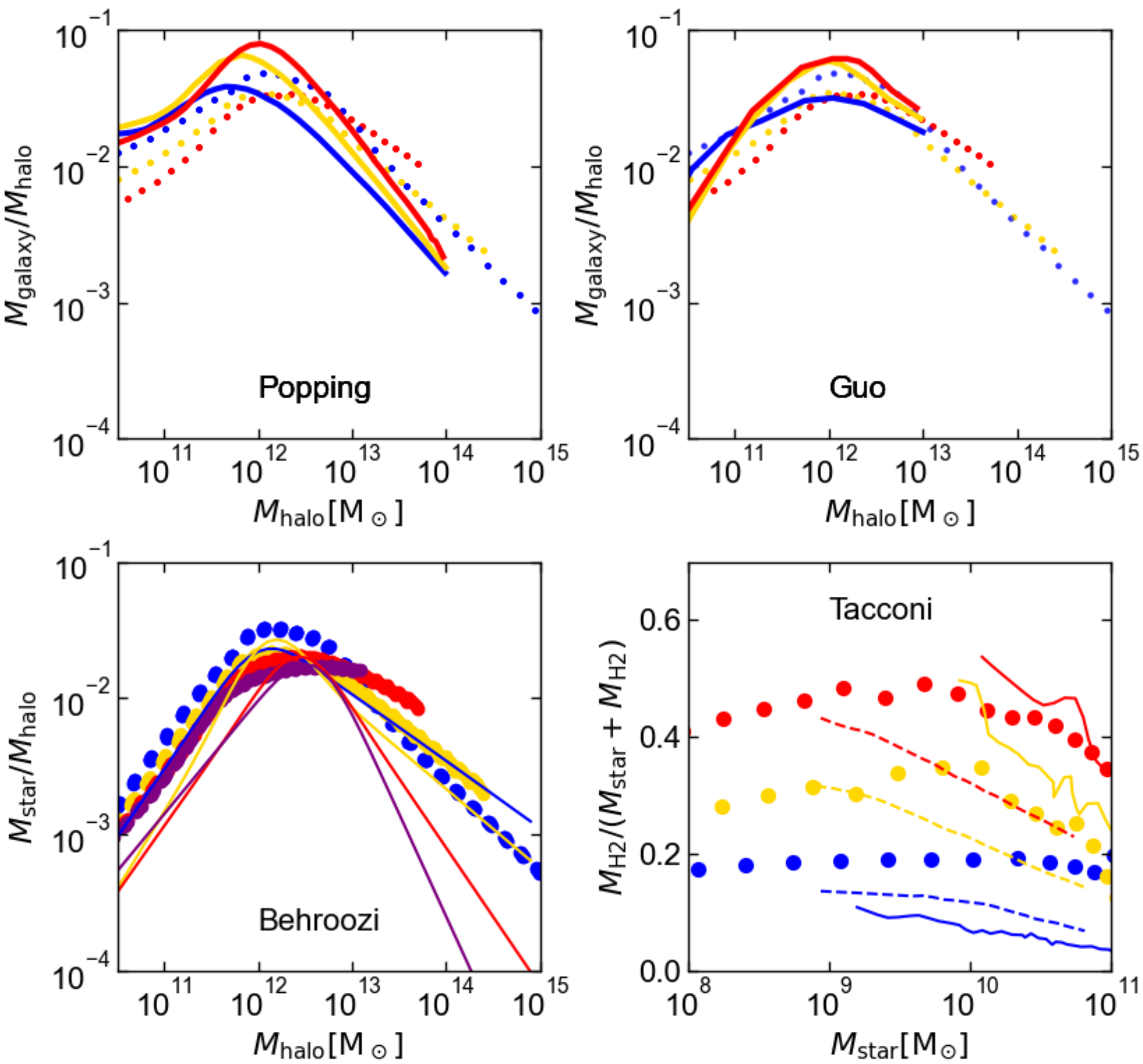

FIG. 5.— Evolution of baryon contents in Model 1. Upper panels: The ratio of the galaxy total mass including ISM and the parent halo mass (dots) compared with the observational results (solid lines) by Popping et al. (2015) (left) and Guo et al. (2023) (right). Lower left: The ratio of the galaxy stellar mass and the parent halo mass (dots) compared with the observational result by Behroozi et al. (2013) (solid lines). Lower right: Mass of molecular hydrogens against the galaxy stellar mass. Dots indicate model results whereas solid lines show the observational data in Tacconi et al. (2020). The semi-analytic model of Popping et al. (2014) is shown by dashed lines. Colors denote redshift: $z$=3 (purple), 2(red), 1(yellow) and 0(blue), respectively, in all panels.

The association between the CGM thermal state and the disk duality advocated in this study is based on circumstantial evidence, namely that it explains the cosmological emergence of dual disks revealed by the observations. It remains to elucidate specific mechanisms that turn the accreted CGM into thick and thin disks. Various mechanisms have been proposed so far regarding the formation of thick disks. Flaring-up of the existing thin disks by mergers and perturbative gravitational forces is one possibility (e.g. Quinn et al. 1993; Xiang & Rix 2022; Yi et al. 2024). The dominance of thick disks in low-mass galaxies may pose a problem for this idea, because those galaxies are expected to have experienced relatively few mergers throughout the cosmic time (e.g. Fakhouri et al. 2010; Zhang et al. 2023). In the case of the MW, the radial migration of kinematically hot stars from inner galactic regions to the solar neighborhood by non-axisymmetric perturbations like bars has been advocated as the origin of its thick disk (Schönrich & Binney 2009). It is, however, demonstrated that the migrant stars actually do not keep their high-dispersion velocities and settle into a configuration that has a similar vertical thickness to the thin disk (Minchev et al. 2013). Accretion of stars from merging satellite galaxies, another candidate process, is not favored as the main formation mechanism in view of the paucity of expected retrograde material (Comerón et al. 2019). Ejection of ISM by supernova feedback (e.g. Dekel & Silk 1986; Mac Low & Ferrara 1999; McQuinn et al. 2019) and the fall back of the ejected gas to galaxy disks (recycling) (e.g. Oppenheimer et al. 2010; Anglés-Alcázar et al. 2017) are important processes in low-mass galaxies. Recent numerical simulations suggest that the ejected gas falls back to galaxies in the cold mode (Decataldo et al. 2024). This process thus may also contribute to the formation of thick disks.

Gravitational instability of galactic disks may provide a promising process. Tsukui et al. (2025) interpreted the emergence of dual-disk structures on the basis of local conditions of gas (in)stability (Toomre (1964)) in galactic disks. Assuming the marginal stability (i.e., $Q \sim 1$) in galactic disks, they deduced disk scale heights from the molecular gas contents observed in nearby and high-redshift galaxies (Tacconi et al. 2020) and concluded that the expected scale heights agree with the observed ones in the JWST sample. In actual disks, the self-regulation of star formation by gravitational instability and star-formation feedback could keep Toomre's Q around unity.

In this instability picture, the relation between dual-disks and clumpy galaxies is an interesting subject. Clump formation in gas-rich disks has been much studied both observationally (e.g Förster Schreiber et al. 2011; Genzel et al. 2011) and theoretically (e.g Noguchi 1998, 1999; Bournaud et al. 2009). These studies have suggested a possibility that thick disks originate in marginally stable clumpy disks. Quite interestingly, the observational analysis of CANDELS galaxies performed by Guo et al. (2015) shows that their sample at high redshift has a high fraction of clumpy disks irrespectively of galaxy masses but the clumpy fraction decreases more rapidly toward low redshift for more massive galaxies. Namely, the clumpy fraction behaves in a similar way to the thick disk fraction across galaxy mass and redshift. Further observational checks of the correlation between thick disks and clumpy disks could help construct a consistent picture of thick-disk formation.

In these two scenarios, the formation of thick disks relies on the gas-richness of galactic disks. The cold accretion of unheated CGM, which proceeds rapidly with the free-fall time of the CGM, is expected to supply the interstellar medium efficiently and make gas-rich disks (e.g Noguchi 2023). It is therefore suggested that the local gravitational instability of galactic disks is the dominant process which links the CGM thermal state and the disk duality. This does not preclude other processes such as listed above operating in transforming the accreted material into thick disks in actual galaxies. Thick disks may thus exhibit diverse properties reflecting their multiple formation routes. Therefore, the connection between the gas accretion modes and two types of disks proposed in this study is a phenomenological interpretation rather than a uniquely established causal picture. Detailed examination of spatial and age structures of galactic disks, with the aid of more advanced model results, will help evaluate the relative importance of the processes discussed here for thick disk formation.

In this study, we tested the models against the observed mass fractions and mean ages of thick and thin disks. The observed disk masses used for comparison have been derived from the photometric decomposition of surface brightness distributions of galaxies. Yet multiple selection methods exist to identify thin and thick disks. In the case of the Milky Way, Alinder et al. (2025) investigated how different classification methods for categorizing stars into the thick and thin disk populations influence the determination of structural properties of the two disks in the Milky Way. They considered two methods that use cuts in the $[\alpha/\mathrm{Fe}] - [\mathrm{Fe/H}]$ and $[\mathrm{Mg/Mn}] - [\mathrm{Al/Fe}]$ plane, the method that uses a dynamical separation in the azimuthal and vertical actions, the method that uses an age-based cut, and the method that uses a kinematic likelihood method. They found that abundance-based approaches offer a clear separation while kinematic and dynamical methods suffer from less clear separations. Large-scale cosmological simulations for galaxy formation provide useful information about chemical and dynamical evolution of different galactic components (e.g. Iza et al. 2025). Especially, the chemical evolution is expected to reflect the gas replenishment processes from halos and therefore the thermal states of the CGM. Future comparison of disk structures of simulated galaxies with their chemical and kinematical properties as done for the Milky Way will provide an important independent test of the present interpretation on the connection of the disk duality and the CGM physical states.

## 7. CONCLUSIONS

In a previous study, we proposed the hypothesis that the duality of galactic disks originates in the thermal state of the circum-galactic medium in dark matter halos on the basis of observations of the Milky Way disk and local disk galaxies. In the recent picture for galaxy formation, the CGM in low-mass galaxies remains cold and accretes onto galaxies on the free-fall timescale. When the halo mass exceeds the critical mass, the CGM is heated by shock waves and AGN feedback to high temperatures and its accretion onto galaxies is slowed. In our hypothesis, this change is interpreted as the transition from thick-disk formation to thin-disk formation. We examined here whether this hypothesis reproduces the properties of dual-disk systems of high-redshift galaxies obtained by the JWST observation. The accretion scheme which reproduced the thick disk fractions in local galaxies was found to produce the mass and redshift dependence of the thick disk mass fraction in broad agreement with the observational data. Earlier transition from single disks to dual disks observed in more massive galaxies is explained by the earlier crossing of the critical mass by more massive halos. The present results reinforce our hypothesis along with its ability to explain the earlier and stronger quenching of star formation observed in more massive galaxies. There are two important caveats. The present models do not include galaxy mergers so that they are not applicable to bulge-dominated galaxies. The suggested link between the CGM accretion modes and disk structures may be indirect and multiple processes may transform accreted material into thick disks in real galaxies.

## DATA AVAILABILITY

The data that support the findings of this study are available from the corresponding author upon reasonable request.

## CODE AVAILABILITY

We have opted not to make available the code used to calculate the evolution of galaxy models because it is a part of the integrated program currently in use for other projects.

## ACKNOWLEDGMENTS

We acknowledge the anonymous referee for detailed and constructive comments which helped improve the manuscript.

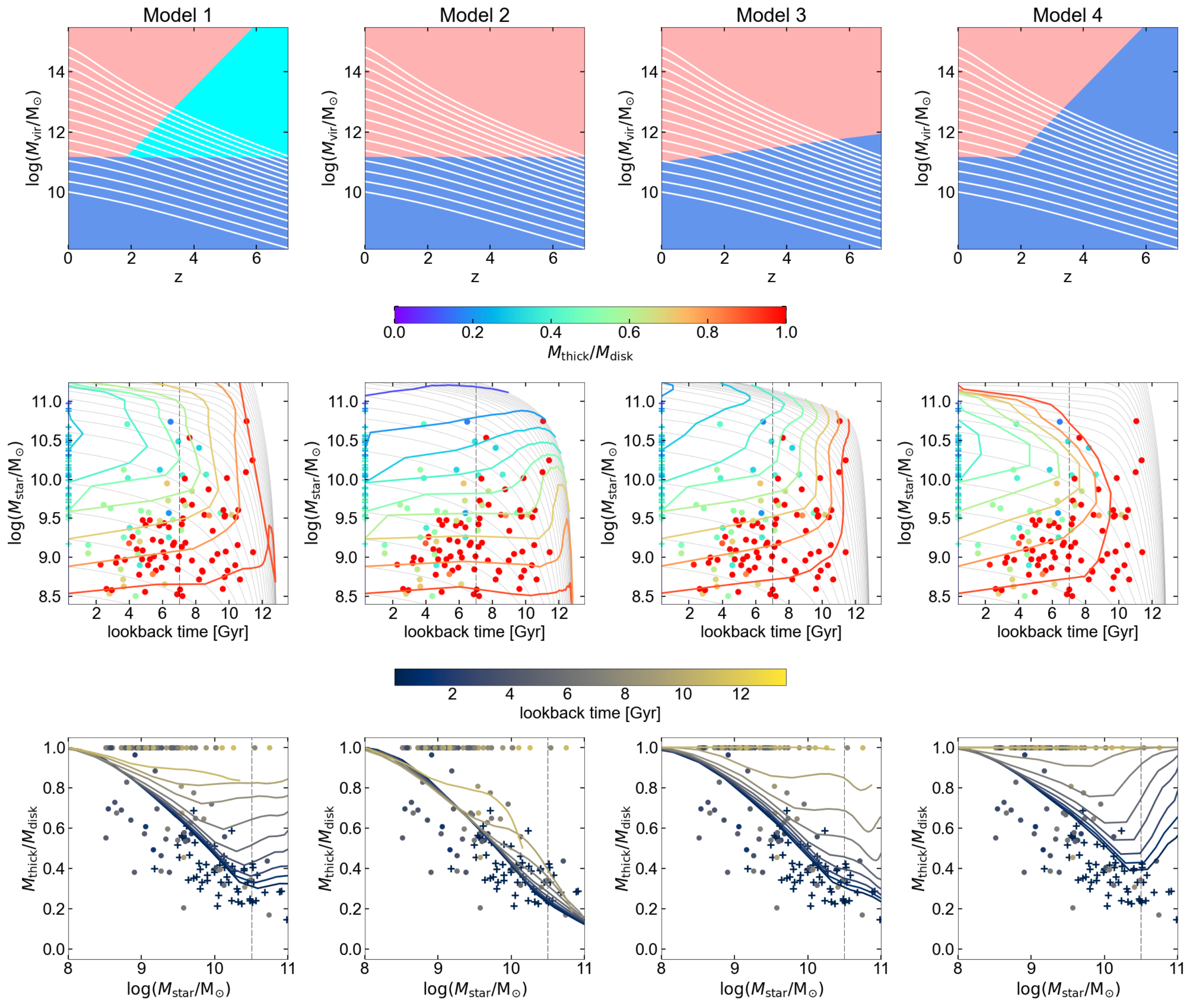

FIG. A.1.— The same as Fig.2, but including all galaxies in Comerón et al. (2014) and Tsukui et al. (2025).

# APPENDIX

## A. EFFECT OF INCLUDING GALAXIES WITH BULGES

As stated in the text, we excluded galaxies with significant bulges listed in Comerón et al. (2014) and Tsukui et al. (2025) from the comparison with model results. Fig.A1 is similar to Fig.2 but includes all the galaxies in both studies. This extension leads mainly to increase in the number of intermediate mass galaxies ($M_{\rm star} \lesssim 10^{9.5}{\rm M}_\odot$) at lookback times $\lesssim 7$ Gyr. These newly-added galaxies tend to possess high thick disk fractions and do not affect the distribution of thick disk fractions on this diagram. We thus consider that the main results obtained in this study are not changed as long as we confine ourselves to galaxies with $B/T \lesssim 0.3$.